\documentclass[twocolumn,aps,prb,showpacs]{revtex4}
\usepackage{mathrsfs}
\usepackage{bm}
\usepackage{multirow}
\usepackage{amsmath,amsfonts,amssymb}
\usepackage{array,booktabs }
\usepackage{graphicx,times,graphics,color,epsfig}
\begin{document}

\title{Enhancement of shot noise due to the fluctuation of Coulomb interaction}
\author{Duo Li, Lei Zhang, Fuming Xu}
\author{Jian Wang}
\email{jianwang@hkusua.hku.hk}

\affiliation{Department of
Physics and the center of theoretical and computational physics, The
University of Hong Kong, Hong Kong, China}
\date{October 28, 2011 under revising}

\begin{abstract}

We have developed a theoretical formalism to investigate the
contribution of fluctuation of Coulomb interaction to the shot noise
based on Keldysh non-equilibrium Green's function method. We have
applied our theory to study the behavior of dc shot noise of atomic junctions using the method
of nonequilibrium Green's function combined with the density
functional theory (NEGF-DFT). In particular, for atomic carbon wire consisting 4
carbon atoms in contact with two Al(100) electrodes, first
principles calculation within NEGF-DFT formalism shows a negative
differential resistance (NDR) region in I-V curve at finite bias due to
the effective band bottom of the Al lead. We have calculated the
shot noise spectrum using the conventional gauge invariant transport
theory with Coulomb interaction considered explicitly on the Hartree
level along with exchange and correlation effect. Although the Fano
factor is enhanced from 0.6 to 0.8 in the NDR region, the expected
super-Poissonian behavior in the NDR region
is not observed. When the fluctuation of Coulomb interaction is
included in the shot noise, our numerical results show that the Fano
factor is greater than one in the NDR region indicating a
super-Poissonian behavior.
\end{abstract}
\maketitle

\section{INTRODUCTION}

Quantum effects have become remarkably significant in nanoscale
semiconductors and the traditional Boltzmann equation is no longer
sufficient to describe transport phenomena. As proposed by Schottky
in his seminar work,\cite{Schottky} partition noise, or shot noise,
is resulted from the quantization of charge. Accordingly, when
electrons are uncorrelated, the classical value of Fano factor,
which describes the magnitude of the electric fluctuation, should be
one. When a Fano factor deviates from one, it shows a signature of
interactions between current flow in different probes. It is known
that shot noise is influenced by two crucial factors, namely, Pauli
principle and Coulomb interaction, which coexist in electronic
systems. Specifically, Pauli interaction can only suppress the Fano
factor below one which corresponds to a sub-Poissonian case and has
been confirmed convincingly by experiments.\cite{Henny, Reznikov,
Kumar, Li} The Coulomb interaction, however, could either reduce the
shot noise, or enhance it so that the Fano factor shows a
super-Poissonian value, depending on details of mesoscopic devices.
Hence, quantum enhancement of shot noise from the classical value
has been the subject of growing interest in recent years and is
explored intensively.\cite{Brown, Liu, Iannaccone, Blanter}

In mesoscopic systems shot noise is very important
since it provides abundant information about transport properties of
conductors, such as kinetics of electrons,\cite{Landauer}
distributions of energy,\cite{Bulashenko} and correlations of
electronic wave functions.\cite{Gramespacher} In addition, people
observed experimentally a shot noise enhancement in the negative differential resistance
(NDR) region with or without the magnetic field in the tunneling structure.\cite{Iannaccone, Kuznetsov}
Various mechanisms have
been proposed to generate a NDR, including enhancement of tunnel
barriers,\cite{Simonian} strong intramolecular
correlations,\cite{Kudasov} band-gap inducement\cite{Rakshit} and so
on. One of early experiments by Li et al suggested that as the NDR
region was approaching, the suppressed value of the Fano factor would
increase.\cite{Li} Further exploration by Iannaccone et al\cite{Iannaccone} showed
that in the NDR region, the shot noise would go through a transition
from a sub-Poissonian value to a super-Poissonian value in a nonlinear fashion.
Nevertheless, a NDR was not a sufficient condition to generate the
enhancement. As Song et al showed that there was no noise
escalation in the super-lattice tunnel diode even though its I-V
curve also exhibited a NDR region.\cite{Song} This led to a conclusion that
charge accumulation, which was related to the internal Coulomb
potential, was ultimately responsible for the super-Poissonian shot
noise. Given the good agreement between numerical calculations from semi-classical theory
and those experiments, the Coulomb interaction was thought to be the
reason for the noise enhancement. Since quantum effect dominates the
transport behavior in mesoscopic systems, a quantum theory of shot noise capable describing
the enhancement in the NDR region is clearly needed. In 1999, Blanter and Buttiker have
studied the shot noise of resonant tunneling quantum well theoretically using scattering matrix method.\cite{Blanter2}
In the nonlinear regime, Coulomb interaction (Hartree level) leads to hysteretic behavior in I-V curve.
By including the fluctuation of Coulomb interaction, they identified an important energy scale, termed
interaction energy, in the Fano factor. They found that in the NDR region where the
interaction energy is very large, a super-poissonian behavior occurs due to the
fluctuation of Coulomb interaction.

Understanding electronic transport properties of atomic-wire based structures
is very important from the scientific viewpoint and due to its potential
applications in molecular electronics. For example, combining the
Lippmann-Schwinger equation and density functional theory (DFT), a NDR in
the tunneling regime of atomic carbon wires was predicted by Lang.\cite{Lang} The shot noise
of silicon atomic wires has also been studied using the same approach\cite{chen}. In this paper,
we develop a general theory for dc shot noise by including the fluctuation of Coulomb interaction. Our
theory is based on non-equilibrium Green's function (NEGF) method which can be coupled with DFT to study transport properties of nano-devices from first principles.
As an application of our theory, we investigate the shot noise of an atomic carbon wire structure with four carbon atoms in the scattering region ($Al-C_4-Al$). Its I-V characteristic and transport
properties have been well understood.\cite{Larade} It was found that a band gap
induced NDR occurs at high bias due to a shift of conduction
channels in the central region. We have used the traditional formula\cite{but1,Wei}
\begin{align}
S_{\alpha \beta} &= (1/2)[<\Delta \hat{I}_\alpha(t) \Delta \hat{I}_\beta(t') >+<\Delta \hat{I}_\beta(t') \Delta \hat{I}_\alpha(t) >] \nonumber\\
&= \frac{q^2}{\pi}\int dE \{[f_\alpha(1-f_\alpha) + f_\beta(1-f_\beta)]Tr[\hat{T}] \nonumber\\
&\quad +(f_\alpha-f_\beta)^2 Tr[(1-\hat{T})\hat{T})]\} \label{eq1}
\end{align}
to calculate the shot noise for $Al-C_4-Al$ structure. Our results show that shot noise is sub-poissonian. When
the fluctuation of Coulomb interaction is included, large shot noise was found in NDR region showing
super-poissonian behavior.
%where $f_{\alpha,\beta}$ are Fermi distribution functions in
%corresponding leads and the transmission coefficient $\hat{T}$ is
%given by $\Gamma_\alpha G^{r} \Gamma_\beta G^{a}$ with $G^{r,a}$
%being the retarded and advanced Green functions of the scattering
%region, respectively.
%Besides, $\Gamma_{\alpha,\beta}$ are the
%linewidth functions related to the coupling of leads and the
%scattering region. This phenomenon was also noticed by
%Nguyen\cite{Do}. However, instead of turning to a semiclassical

Our paper is organized as follows. In section II, we derive a general theory
for dc shot noise when the fluctuation of Coulomb interaction is included
in the first order. The detailed derivation is given in Appendix. In
section III, we describe some technical details and show the numerical
results in the atomic carbon chain system along with an analysis and
discussion of the result. Finally, the summary is given in section IV.

\section{THEORETICAL FORMALISM}
In this section, a NEGF theory is developed to calculate dc shot
noise in the regime of NDR, which
involves the Coulomb interaction between electrons. A key ingredient
of the new theory is that to account for large shot noise in the NDR
region both self-consistent Coulomb potential and its fluctuation
have to be considered.

\subsection{General Expression}

We start from a quantum coherent two-lead conductor defined by the Hamiltonian
\begin{align}
\hat{H}_0 &= \sum_{k\alpha} \epsilon_{k\alpha} \hat{C}^{\dagger}_{k\alpha}\hat{C}_{k\alpha} + \sum_n (\epsilon_n +qU_n) d^\dagger_n d_n\nonumber\\
&\quad +\sum_{k\alpha n}[t_{k\alpha n} \hat{C}^{\dagger}_{k\alpha} \hat{d}_n + c.c.]
\end{align}
where $\hat{C}^\dagger_{k\alpha}(\hat{C}_{k\alpha})$,
$d^\dagger_n(d_n)$ are the creation (annihilation) operators of
electrons in leads and the scattering region, respectively. The
first term describes leads that dc voltages are applied on, and
$\epsilon_{k\alpha}=\epsilon^{(0)}_{k\alpha} +qv_\alpha$ which
$\epsilon^{(0)}_{k\alpha}$ is energy levels in the lead $\alpha$ and
$v_\alpha$ stands for an external voltage. The second term is for
the isolated central region, where the self-consistent internal
Coulomb potential under the Hartree approximation is defined as
\begin{equation}
U_n = \sum_m V_{nm} <d^\dagger_{m} d_{m}> \label{coulomb}
\end{equation}
where $V_{nm}$ is a matrix element of the Coulomb potential. In the
real space $V(x,x') = 1/|x-x'|$, $q$ is the electron charge. The
last term corresponds to a coupling between the central region and
leads described by a coupling constant $t_{k\alpha n}$.

The current operator of the lead $\alpha$ is defined as ($\hbar=1$):
\begin{align}
\hat{I}_{\alpha_0}(t) = q\frac{d\hat{N}_\alpha}{dt}
\end{align}
where $\hat{N}_\alpha=\sum_k \hat{C}^\dagger_{k\alpha}
\hat{C}_{k\alpha}$ is the number operator for electrons in the lead
$\alpha$.

From the Heisenberg equation of motion,
\begin{eqnarray}
\frac{d\hat{N}_\alpha}{dt} = -i[\hat{N}_\alpha, \hat{H}_0]
\end{eqnarray}
we have
\begin{eqnarray}
\frac{d\hat{N}_\alpha}{dt} =-i\sum_{kn}[t_{k\alpha n}
\hat{C}^\dagger_{k\alpha} \hat{d}_n] + h.c.
\end{eqnarray}
where $\hat{H}$ is the system Hamiltonian with constant Coulomb
potential and h.c. denotes the Hermitian conjugate.

Hence the current operator becomes
\begin{equation}
\hat{I}_{\alpha_0}(t) = -iq\sum_{kn}[t_{k\alpha n} \hat{C}^\dagger_{k\alpha}(t) \hat{d}_n(t)] + h.c.
\label{current}
\end{equation}

On the mean field level, the current is a functional of Coulomb
interaction, i.e., $\hat{I} = \hat{I}[\langle \hat{U} \rangle]$.
Here we have treated the operator of Coulomb interaction ${\hat U}$
as a C number meaning that the fluctuation of Coulomb interaction is
assumed not important. However, this is not always true. For
instance, in order to reflect the Coulomb interaction between
electrons in the NDR region, we have to consider the fluctuation of
the Coulomb potential. Fig.1 shows the physical picture of the NDR.
For simplicity, we assume that the scattering region has one
resonant level $E_0$ with a width characterizing the lifetime of the
resonant level. We also assume that there is an effective band
bottom for the lead which is crucial for the phenomenon of NDR. As
shown in Fig.1, when the bias voltage is increased the current
increases because the resonant level is brought down by the external
bias. As the bias is increased further such that the resonant level
falls below the band bottom of the lead the current starts to
decrease giving rise to the NDR. The above physical picture is
static where the Coulomb potential is included on the mean field
level and the correlation effect of Coulomb interaction has been
neglected. For the current correlation in the NDR region, the
correlation effect of Coulomb interaction has to be
considered. In this picture, when the resonant
energy level $E_0$ is about to fall below the band bottom of the
lead, the internal potential of the scattering region due to the
Coulomb interaction of injected electron will push it up, leading to
a positive correlation between incoming electron
flows.\cite{Iannaccone} This positive correlation is a dynamic
process and can not be described by a Hartree field. In another
word, the fluctuation of Coulomb interaction has to be considered
for the positive correlation in the NDR region. As demonstrated by
Larade\cite{Larade} and confirmed by our calculation, for the atomic
wire with even number carbon atoms like $C_4$ and $C_6$, there is an
effective band bottom responsible for the NDR. Therefore, the
fluctuation of Coulomb potential $\hat{U}$ should be important. For
odd number wires such as $C_5$ and $C_7$, however, there is no
apparent NDR effect. Hence there no effective band bottom and the
fluctuation of $\hat{U}$ can be neglected.

To treat the fluctuation of Coulomb interaction, we follow the idea
of Ref.\onlinecite{Blanter2} and expand the current in terms of
Coulomb potential operator about its equilibrium value up to linear
order. After the expansion, the total current in the real space
could be expressed as
\begin{align}
\hat{I}_\alpha(t) &\simeq \hat{I}_{\alpha_0}(t) + \sum_i \frac{\delta{\hat{I}_{\alpha_0}(t)}}{\delta{\hat{U}_i}(t)}
\big|_{\hat{U}_i(t) = U_i} (\hat{U_i}(t) - U_i) \nonumber\\
                  &\simeq \hat{I}_{\alpha_0}(t) + \sum_i \frac{\delta{I_{\alpha}}}{\delta{U_i}}(\hat{U_i}(t) - U_i) \nonumber\\
                  &= \hat{I}_{\alpha_0}(t) + \sum_i \lambda_{\alpha i} \delta\hat{U}_i(t)
\end{align}
where $I_{\alpha} = <\hat{I}_{\alpha_0}>$, $U_i = <\hat{U}_i>$, and
$\delta\hat{U}_i(t) = \hat{U}_i(t) - U_i$.  We have also introduced
a quantity $\lambda_{\alpha i}=\delta{I_{\alpha}}/\delta{U_i}$.

\begin{figure}
\includegraphics[width=8cm,totalheight=6cm,angle=0]{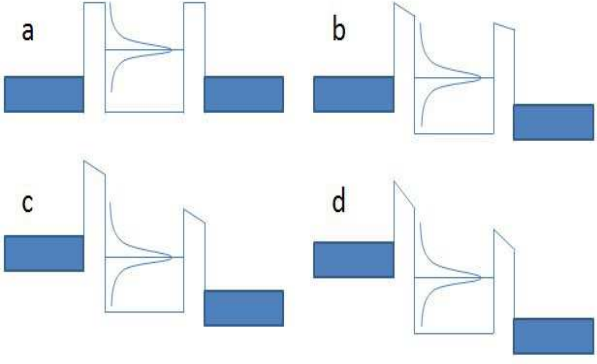}
\caption{Graphical band profile for nanoscale devices in the presence of bias. At the zero bias, the resonant level is above the Fermi level (a). As the bias increases, the Fermi level of left lead increases and is above the resonant level (b) and (c), giving rise to a large increase of current. As the bias increases further, the resonant level is below the band bottom of the left lead (d) leading to a NDR.}
\label{Fig.1}
\end{figure}

It is easy to see that this new current operator gives the same
current but different shot noise. With the new current operator, the
current correlation is obtained up to the linear order in $\delta
U$,
\begin{eqnarray}
S^{(1)}_{\alpha \beta} &&= <\Delta\hat{I}_\alpha(t)\Delta\hat{I}_\beta(t')> = <(\hat{I}_\alpha(t)\hat{I}_\beta(t')>-I_\alpha I_\beta  \nonumber \\
&&\simeq <\hat{I}_{\alpha_0}(t)\hat{I}_{\beta_0}(t')> -
I_\alpha I_\beta + \Delta^{(1)}_{\alpha\beta}. \label{eq9}
\end{eqnarray}
where $S_{\alpha \beta}=(1/2)[S^{(1)}_{\alpha \beta}+S^{(2)}_{\alpha \beta}]$.
The first two terms in Eq.(\ref{eq9}) corresponds to the current correlation
$S^{(1)}_{\alpha \beta 0}$ in the absence of Coulomb potential
fluctuation. They have been calculated before and is given by
Eq.(\ref{eq1}). The last term in Eq.(\ref{eq9}) is defined as
\begin{align}
\Delta^{(1)}_{\alpha\beta} &= \sum_i [\lambda_{\beta i} <\hat{I}_{\alpha_0}(t) \hat{U}_i(t')> +
\lambda_{\alpha i}< \hat{U}_i(t) \hat{I}_{\beta_0}(t')>  \nonumber \\
&-(\lambda_{\alpha i} U_i I_{\beta} +\lambda_{\beta i} U_i I_{\alpha})] \label{eq10}
\end{align}
Using the NEGF method, we have derived the expression of
$\Delta^{(1)}_{\alpha\beta}$ (see the appendix). Finally, the shot noise
in the presence of Coulomb potential fluctuation is written as
\begin{align}
S_{\alpha \beta} &= S_{\alpha \beta 0} + \Delta_{\alpha\beta} \label{eq11}
\end{align}
where
\begin{align}
\Delta_{\alpha\beta}&= -\frac{q^2}{2\pi} \sum_{ij} [\lambda_{\beta
i} V_{ij} {\rm Im}(\Xi_{\alpha jj}) + \lambda_{\alpha i} V_{ij} {\rm Im}(\Xi_{\beta
jj})] \label{eq12}
\end{align}
with
\begin{align}
\lambda_{\alpha i} = -\frac{q}{2\pi} \int dE \sum_\beta
(f_\alpha-f_\beta) (G^r \Gamma_\beta G^a \Gamma_\alpha
G^r)_{ii}+h.c. \label{eq13}
\end{align}
and
\begin{align}
\Xi_\alpha &= i\int dE [G^r \Gamma_\alpha G^r \Sigma^> G^a f_\alpha - G^r
\Sigma^< G^a \Gamma_\alpha G^a (f_\alpha-1)\nonumber\\
           &\quad + G^r \Sigma^< G^a \Gamma_\alpha G^r \Sigma^> G^a] \label{eq14}
\end{align}
Eqs.(\ref{eq11})-(\ref{eq14}) is the central result of this paper. To calculate the shot
noise, one has to solve the Green's function together with the
Poisson equation,
\begin{eqnarray}
G^r = \frac{1}{E-H-U-V_{xc}-\Sigma^r}
\end{eqnarray}
and
\begin{eqnarray}
\nabla^2 U(x) = 4\pi i q \int \frac{dE}{2\pi} G^<(E,x,x) \label{eq16}
\end{eqnarray}
where $V_{xc}$ is the potential due to the exchange and correlation
effect in the first principles calculation.

For a two-terminal device at zero temperature, we set $\alpha=L,
\beta=R$ and $v_R > v_L=0$ so that $f_L (f_L-1)=0$ and
$f_R (f_L-1)=0$ at zero temperature. We then obtain
\begin{align}
&\quad \Xi_L = i\int dE [G^r \Gamma_L G^r \Sigma^> G^a f_L \nonumber\\
&- G^r \Sigma^< G^a \Gamma_L G^a (f_L-1) + G^r \Sigma^< G^a \Gamma_L G^r \Sigma^> G^a] \nonumber\\
&= \int_{E_F-qv_R}^{E_F} dE (G^r \Gamma_L G^r \Gamma_R G^a + i G^r
\Gamma_L G^a \Gamma_L G^r \Gamma_R G^a) \label{eq17}
\end{align}

Similarly, we have
\begin{align}
\Xi_R = \int_{E_F-qv_R}^{E_F} dE (G^r \Gamma_R G^r \Gamma_L G^a + i
G^r \Gamma_R G^a \Gamma_R G^r \Gamma_L G^a) \label{eq18}
\end{align}

In addition, from Eq.(\ref{eq13}), we have
\begin{align}
\lambda_L &= -\frac{q}{2\pi} \int_{E_F-qv_R}^{E_F} dE (G^r \Gamma_R G^a \Gamma_L G^r) + h.c. \nonumber \\
\lambda_R &= \frac{q}{2\pi} \int_{E_F-qv_R}^{E_F} dE (G^r \Gamma_L
G^a \Gamma_R G^r )+ h.c. \label{eq19}
\end{align}
With Eqs.(\ref{eq17}), (\ref{eq18}), and (\ref{eq19}), the two-probe
shot noise can be calculated.

\section{NUMERICAL RESULTS}

\begin{figure}
\includegraphics[width=9cm,totalheight=6cm,angle=0]{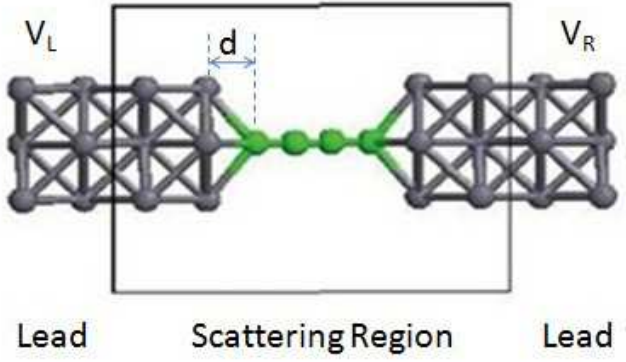}
\caption{The schematic plot of $Al-C_4-Al$ system. The atomic wire
with four carbon atoms is linked by two semi-infinite Al electrodes.
The (100) direction Al electrodes extend to $\pm \infty$ where
electron reservoirs are located.} \label{Fig.2}
\end{figure}

In this section we use state-of-the-art first-principle quantum
transport package MATDCAL to investigate the general transport
properties of atomic carbon-chain systems coupled with Al leads. In
the package DFT is carried out within the formalism of the Keldysh
nonequilibrium Green's function. Numerically, the effective
Kohn-Sham (KS) equations is solved by a linear combination of the
atomic orbitals (LCAO) basis set. We define the atomic core by a
nonlocal norm conserving pseudopotential and treat the
exchange-correlation at the LDA level. DFT determines the atomic
structure and the system Hamiltonian while NEGF contributes to the
nonequilibrium transport properties. Under an external bias the
transport boundary conditions are treated by the real space
numerical techniques. For further references, the theoretical
background and practical execution of this formalism can be found in
Ref.\onlinecite{Taylor}. A numerical error tolerance is set to be
$10^{-4}$ to confirm self-consistency.

Generally speaking, we have carried out our calculation on the atomic
chain structure with four carbon atoms $Al-C_4-Al$.\cite{Larade} The
carbon chain lies in the central simulation box in contact with
electron reservoirs through two semi-infinite Al electrodes. The
schematic structure is shown in Fig.2 where there are 18 Al atoms in
the unit cell of the semi-infinite electrodes with a cross section
along (100) direction. The contact distance between the Al electrode
and the carbon chain is fixed at 0.378 a.u. while the distance
between the nearby carbon atoms is equal to 2.50 a.u. In our
calculation, we have set temperature to be zero.

\begin{figure}
\includegraphics[width=9cm,totalheight=6cm,angle=0]{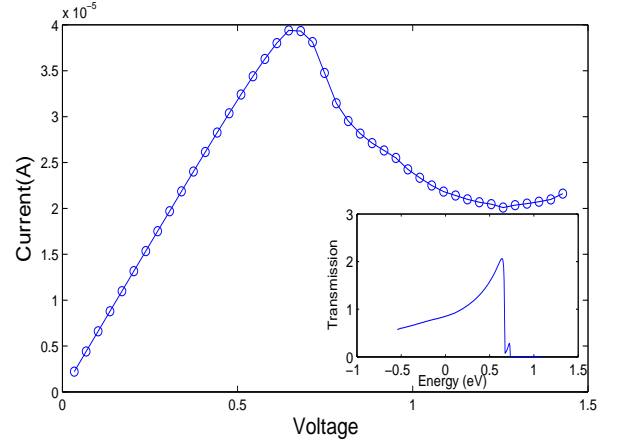}
\caption{The I-V curve of $Al-C_4-Al$ structure. A NDR region begins
to show up around 0.65 V. Inset: the transmission coefficient
for this system at zero bias.} \label{Fig.3}
\end{figure}

Technically, the correction term in Eq.(\ref{eq12}) has to be
solved in real space due to the Coulomb like interaction involving
$V_{ij}$ which reads $V(x,x') = 1/|x-x'|$ in real space. Since
quantities $\lambda$ and $\Xi$ play the role of charge, we can define
the potential induced by $\Xi$,
\begin{align}
\Omega_{\alpha x} = \sum_{x'} V(x,x') {\rm Im}(\Xi_{\alpha}( x',x'))
\end{align}
Since $\nabla^2_x V(x,x')=-4\pi \delta(x-x')$, $\Omega_{\alpha}$
satisfies the Poisson like equation,
\begin{align}
\nabla^2 \Omega_{\alpha}(x) = -4\pi {\rm Im}(\Xi_{\alpha} (x,x))
\end{align}
We solve this equation for the leads to obtain the boundary condition for the scattering region.
It turns out $\Omega_{\alpha}(x)$ in the lead is very small so that the boundary condition
of $\Omega_{\alpha}(x)$ can be safely set to zero. Once $\Omega_{\alpha}(x)$ is obtained the
correction term can be easily calculated from Eq.(\ref{eq12})
\begin{align}
\Delta_{\alpha\beta} = -\frac{q^2}{2\pi} \int [\lambda_\beta(x) \Omega_{\alpha}(x)
+ \lambda_\alpha(x) \Omega_\beta(x)] dx
\end{align}

The I-V characteristics is shown in Fig.3, where the inset plots the
transmission coefficient T versus the energy E at zero bias. The
shot noise and the corresponding Fano factor are shown in Fig.4. In Fig.3 and 4,
the Coulomb interaction is included on the Hartree level and the
Coulomb potential fluctuation is neglected. Following observations
are in order. (1) The I-V curve is similar to the result obtained
by Larade et al.\cite{Larade} (2) At zero bias the resonant energy
is higher than Fermi energy of the system (chosen to be zero). As we
apply a voltage to the right electrode, the resonant energy level
should drop and move closer to Fermi energy of the left electrode.
With the increasing of voltage, the effective band bottom of
the emitter and the main resonant level would be aligned, and this
gives the maximum current around 0.65 V. When the voltage increases
further, a significant decrease of current occurs.
(3) The shot noise in the absence of Coulomb potential fluctuation
has a similar behavior as that of I-V except
that the maximum is at 0.7V instead of 0.65V. (4) The Fano factor
is nearly a constant of order of 0.6 in positive differential
resistance (PDR) region when bias is smaller than 0.6V. It starts to
increase sharply upon entering the NDR region and the Fano factor
eventually shoots up to 0.8. We conclude that although the Fano
factor calculated on the Hartree level shows enhancement in the NDR
region, it is still sub-Poissonian which does not agree with
experimental result.\cite{Iannaccone}

\begin{figure}
\includegraphics[width=9cm,totalheight=6cm,angle=0]{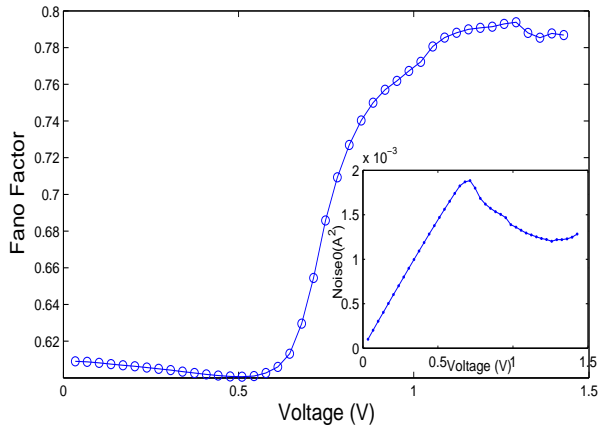}
\caption{The Fano factor derived from the unmodified model
which shows sub-Poissonian behavior. Inset: the noise spectrum calculated
by Eq.(\ref{eq1})} \label{Fig.4}
\end{figure}

In general, when electrons tunnel through the left barrier to occupy empty
energy levels, Pauli principle inhibits other tunneling electrons
to reach the same energy level but higher ones. As a consequence Pauli exclusion
principle gives the negative effect for current correlation. Coulomb interaction,
however, can give either positive or negative correlation effect to shot noise. This
can be understood as follows. It is known that the maximum current corresponds to the
situation that the energy of incoming electron is in line with the resonant level.
Hence in both PDR and NDR regions, most of electrons are off resonance.
Due to the Coulomb interaction the incoming electron can push up the resonant level
so that the next electron will be further away from the resonance in the PDR region or closer
to the resonance in the NDR region, giving rise negative or positive correlations.
Our numerical results indeed confirm this physical picture.
In Fig.5 and 6 we present the result of shot noise and Fano factor in the presence
of fluctuation of Coulomb interaction. We have also included the shot noise and
Fano factor in the absence of Coulomb potential fluctuation for comparison.
In Fig.5, we see that the correction term solved via the Poisson like equation is
very small at low bias when $V<0.5$ and becomes negative until around 1.0V where shot noise
increases sharply. In Fig.6, we see that a large Fano factor great than 3 occurs near
$V=1.0$. This result is in qualitatively agreement with others' work.\cite{Blanter2,Do,Iannaccone}
\begin{figure}
\includegraphics[width=9cm,totalheight=6cm,angle=0]{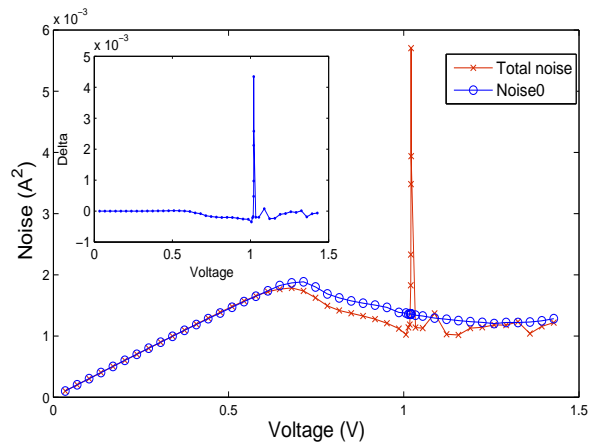}
\caption{The noise spectrum plotted as a function of the applied
voltage at 0 K. Inset: the correction term for a two-terminal system.} \label{Fig.5}
\end{figure}

\begin{figure}
\includegraphics[width=9cm,totalheight=6cm,angle=0]{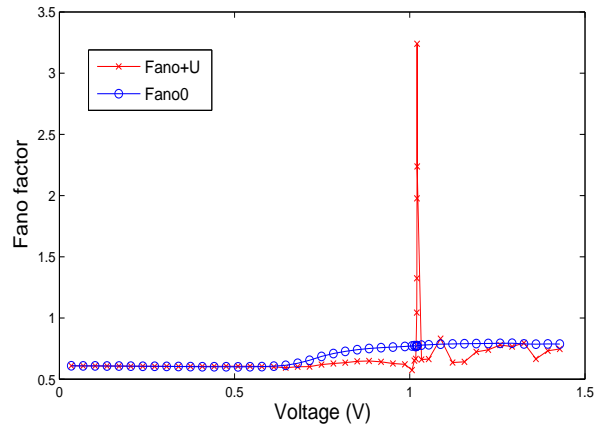}
\caption{The Fano factor plotted as a function of the applied
voltage for $Al-C_4-Al$ system at 0 K. Circle: the Fano factor is
calculated by the Eq.(\ref{eq1}) for the same system.
}
\label{Fig.6}
\end{figure}

\section{CONCLUSION}

The traditional formalism can only describe the suppression of shot
noise, which corresponds to the PDR region. In order to treat enhancement
of shot noise correctly in the NDR region, we have to include the
fluctuation of Coulomb interaction. In this paper we have developed such
a general dc theory for calculating the shot-noise in the NDR region.
The theoretical framework is based on the combination of NEGF-DFT
formalism with the self-consistent Coulomb potential and its fluctuation
included. Our theory (Eq.(\ref{eq11})) can also be applied to mesoscopic
conductors. We have applied our theory to molecular devices. Specifically, we
have calculated the shot noise of $Al-C_4-Al$ structure which is an ideal
system since its I-V curve exhibits a NDR region. We found a large Fano factor
in the NDR region exhibiting super-Poissonian behavior.

\section{Appendix}
In this appendix, we will derive the expression of
$\lambda_{\alpha,\beta}$ and $\Delta_{\alpha\beta}$ using the theory
of NEGF.

\bigskip
\noindent {\bf Expression of $\lambda_{\alpha,\beta}$}
\bigskip

Using the NEGF, the current is given by
\begin{align}
I_{\alpha} = -\frac{q}{2\pi}\int dE \sum_\beta[Tr(\Gamma_\alpha G^r \Gamma_\beta G^a)](f_\alpha-f_\beta)
\end{align}
where $f_{\alpha,\beta}$ are Fermi distribution functions in
corresponding leads, $G^{r,a}$ are the respective retarded and
advanced Green functions of the scattering region and
$\Gamma_{\alpha,\beta}$ are the linewidth functions related to the
coupling of leads and the scattering region.

To obtain $\lambda_{\alpha}$, we first calculate the following quantity,
\begin{align}
\lambda_{\alpha i} (E)&=  -\frac{q}{2\pi}\sum_\beta Tr \frac{\partial}{\partial U_{i}}[\Gamma_\alpha G^r \Gamma_\beta G^a] \nonumber\\
&= -\frac{q}{2\pi} \sum_\beta Tr [\Gamma_\alpha (\frac{\partial G^r}{\partial U_{i}})
\Gamma_\beta G^a + \Gamma_\alpha G^r \Gamma_\beta (\frac{\partial G^a}{\partial U_{i}})] \nonumber \\
&= -\frac{q}{2\pi} \sum_\beta Tr [\Gamma_\alpha (\frac{\partial G^r}{\partial U_{i}}) \Gamma_\beta G^a]+h.c. \label{eq21}
\end{align}

The quantity $\partial_U G^r$ can be calculated from the Dyson equation\cite{Haug}
\begin{align}
G^r = G^r_0+G^r_0 U G^r
\end{align}
where $G^r_0$ is the retarded Green's function in the absence of
Coulomb interaction. Taking the derivative with respect to $U_i$, we
have
\begin{align}
\frac{\partial}{\partial U_{i}}G^r = \frac{\partial}{\partial U_{i}}(G_0^r U G^r)
= G_0^r D_{i} G^r + G_0^r U \frac{\partial G^r}{\partial U_{i}} \label{eq23}
\end{align}
where $D_i$ is a diagonal matrix with the matrix element $(D_i)_{jk}
= \delta_{ji} \delta_{ki}$, i.e., there is only one nonzero matrix
element. From Eq.(\ref{eq23}), we find
\begin{align}
\frac{\partial}{\partial U_{i}}G^r = \frac{1}{1 - G_0^r U} G_0^r D{i} G^r
=G^r D_{i} G^r \label{eq24}
\end{align}
where we have used the following Dyson equation again,
\begin{align}
G^r &= \frac{1}{1 - G_0^r U} G_0^r \nonumber
\end{align}
Substituting Eq.(\ref{eq24}) into Eq.(\ref{eq21}), we can get
\begin{align}
\lambda_{\alpha i} (E) &= -\frac{q}{2\pi}\sum_\beta Tr(\Gamma_\alpha G^r D_{i} G^r \Gamma_\beta G^a) +h.c. \nonumber\\
&= -\frac{q}{2\pi}\sum_\beta Tr( G^r \Gamma_\beta G^a \Gamma_\alpha G^r D_{i})  +h.c. \nonumber\\
&= -\frac{q}{2\pi}\sum_\beta (G^r \Gamma_\beta G^a \Gamma_\alpha G^r)_{ii} + h.c.
\end{align}
which is equivalent to Eq.(\ref{eq13}).

\bigskip
\noindent {\bf Expression of $\Delta_{\alpha\beta}$}
\bigskip

For simplicity, we only deal with the first term of
$\Delta_{\alpha\beta}$ explicitly, i.e. $ <\hat{I}_{\alpha_0}(t)
\hat{U}_i(t')>$. The second term can be calculated similarly. Using
the current and Coulomb potential operators in Eqs.(\ref{current})
and (\ref{coulomb}), we obtain
\begin{align}
& <-iq^2\sum_{kn}[t_{k\alpha n} \hat{C}^\dagger_{k\alpha}(t) \hat{d}_n(t) -h.c.]
\sum_m V_{im} \hat{d}_m^\dagger(t') \hat{d}_m(t')> \nonumber\\
&= -iq^2 \sum_{knm} V_{im} [t_{k\alpha n}<\hat{C}^\dagger_{k\alpha}(t) \hat{d}_n(t) \hat{d}_m^\dagger(t') \hat{d}_m(t')> \nonumber\\
&\quad - t_{k\alpha n}^*<\hat{d}_n^\dagger(t) \hat{C}_{k\alpha}(t) \hat{d}_m^\dagger(t') \hat{d}_m(t')>] \nonumber\\
&= -iq^2 \sum_{knm} V_{im} [t_{k\alpha n}<\hat{C}^\dagger_{k\alpha}(t) \hat{d}_m(t')> <\hat{d}_n(t) \hat{d}_m^\dagger(t')> \nonumber\\
&\quad -t_{k\alpha n}^* <\hat{d}_n^\dagger(t)  \hat{d}_m(t')> <\hat{C}_{k\alpha}(t) \hat{d}_m^\dagger(t')>] +I_\alpha U_i
\end{align}

In terms of Green's functions\cite{Datta},
\begin{align}
G_{m,k\alpha}^<(t,t') &= i <\hat{C}^\dagger_{k\alpha}(t') \hat{d}_m(t)>  \\
G_{nm}^>(t',t) &= -i <\hat{d}_n(t') \hat{d}_m^\dagger(t)>  \\
G_{mn}^<(t,t') &= i <\hat{d}^\dagger_n(t') \hat{d}_m(t)>  \\
G_{k\alpha,m}^>(t',t) &= -i <\hat{C}_{k\alpha}(t') \hat{d}_m^\dagger(t)>
\end{align}
we obtain,
\begin{align}
&<\hat{I}_{\alpha_0}(t) \hat{U}_i(t')> \nonumber\\
&= -iq^2 \sum_{knm} V_{im} [t_{k\alpha n}G_{m,k\alpha}^<(t,t') G_{nm}^>(t',t) \nonumber\\
&\quad - t_{k\alpha n}^* G_{mn}^<(t,t') G_{k\alpha,m}^>(t',t)] + I_\alpha U_i
\end{align}
Applying the Langreth theorem of analytic
continuation\cite{Langreth} and suppressing time indices, we have
\begin{align}
G_{m,k\alpha}^< &= \sum_l (G_{ml}^r t_{k\alpha l}^* g_{k\alpha}^< + G_{ml}^< t_{k\alpha l}^* g_{k\alpha}^a) \\
G_{k\alpha,m}^> &= \sum_l (g_{k\alpha}^> t_{k\alpha l} G_{lm}^a + g_{k\alpha}^r t_{k\alpha l} G_{lm}^>)
\end{align}
where $g_{k\alpha}^{<,>,r,a}$ are the corresponding Green's
functions in the lead $\alpha$.

For dc transport, the Green functions depend only on $t'-t$. After
the Fourier transform from time to energy, it becomes
\begin{align}
& <\hat{I}_{\alpha_0} \hat{U}_i> - I_\alpha U_i\nonumber\\
&= -\frac{iq^2}{2\pi} \sum_{knml} V_{im} \int dE [t_{k\alpha n}(G_{ml}^r t_{k\alpha l}^* g_{k\alpha}^< \nonumber\\
&\quad + G_{ml}^< t_{k\alpha l}^* g_{k\alpha}^a)G_{nm}^> - t_{k\alpha n}^* G_{mn}^< (g_{k\alpha}^> t_{k\alpha l} G_{lm}^a \nonumber\\
&\quad + g_{k\alpha}^r t_{k\alpha l} G_{lm}^>)] \nonumber\\
&= -\frac{iq^2}{2\pi} \sum_{m} V_{im} \int dE [(G^r \Sigma_{\alpha }^< + G^< \Sigma_{\alpha }^a)G^>- G^< (\Sigma_{\alpha}^> G^a \nonumber\\
&\quad + \Sigma_{\alpha}^r G^> )]_{mm} \nonumber \\
&=-\frac{iq^2}{2\pi} \sum_{m} V_{im} [\Xi_\alpha]_{mm} \label{eq34}
\end{align}
where $\Sigma^{<,>,r,a}_\alpha$ are the corresponding functions of
self-energy due to the lead $\alpha$. Using the Keldysh equation and
properties of Green's functions\cite{Haug},
\begin{align}
G^< &= G^r\Sigma^< G^a  \\
\Sigma_\alpha^a(E) - \Sigma_\alpha^r(E) &= i \Gamma_\alpha(E-qv_\alpha) \\
\Sigma_\alpha^<(E) &= i \Gamma_\alpha(E-qv_\alpha) f_\alpha(E)  \\
\Sigma_\alpha^>(E) &= i \Gamma_\alpha(E-qv_\alpha) (f_\alpha(E)-1)
\end{align}
we find
\begin{align}
\Xi_\alpha &= i\int dE [G^r \Gamma_\alpha G^r \Sigma^> G^a f_\alpha - G^r \Sigma^< G^a \Gamma_\alpha G^a (f_\alpha-1)\nonumber\\
           &\quad + G^r \Sigma^< G^a \Gamma_\alpha G^r \Sigma^> G^a]
\end{align}
Substituting this equation into Eqs.(\ref{eq34}) and (\ref{eq10}), we
finally arrive at
\begin{align}
\Delta^{(1)}_{\alpha\beta}&= -\frac{iq^2}{2\pi} \sum_{ij} (\lambda_{\beta i} V_{ij} \Xi_{\alpha j} - \lambda_{\alpha i} V_{ij} \Xi_{\beta j}^*) \label{del1}
\end{align}
where we have used the fact that $<\hat{I}_{\alpha_0}(t) \hat{U}_i(t')> =< \hat{U}_i(t) \hat{I}_{\alpha_0}(t')>^\dagger$.

\section{ACKNOWLEDGEMENTS}
We thank Dr. Y.X. Xing for checking all the algebras in the theory part.
This work was financially supported by Research Grant Council
(HKU 705409P) and University Grant Council (Contract No. AoE/P-04/08) of the Government of HKSAR.


\begin{thebibliography}{00}

\bibitem{Schottky}
W. Schottky, Ann. Phys.(Leipzig) {\bf 57}, 541 (1918).

\bibitem{Henny}
M. Henny, S. Oberholzer, C. Struck, and C. Sch\"onenberger, Phys. Rev. B {\bf 59}, 2871 (1999).

\bibitem{Reznikov}
M. Reznikov, M. Heiblum, H. Shtrikman, and D. Mahalu, Phys. Rev. Lett. {\bf 75}, 3340 (1995).

\bibitem{Kumar}
A. Kumar, L. Saminadayar, D. C. Glattli, Y. Jin, and B. Etienne, Phys. Rev. Lett. {\bf 76}, 2778 (1996).

\bibitem{Li}
Y. P. Li, D. C. Tsui, J. J. Heremans, J. A. Simmons, and G. W. Weimann, Appl. Phys. Lett. {\bf 57}, 774 (1990).

\bibitem{Brown}
E. R. Brown, IEEE Trans. Electron Devices {\bf 39}, 2686 (1992).

\bibitem{Liu}
R. C. Liu, B. Odom, Y. Yamamoto and S. Tarucha, Nature (London) {\bf 391}, 263 (1998).

\bibitem{Iannaccone}
G. Iannaccone, G. Lombardi, M. Macucci, and B. Pellegrini, Phys. Rev. Lett. {\bf 80}, 1054 (1998).

\bibitem{Blanter}
Y. M. Blanter and M. B\"uttiker, Phys. Rep. {\bf 336}, 1 (2000).

\bibitem{Landauer}
R. Landauer, Nature (London) {\bf 392}, 659 (1998).

\bibitem{Bulashenko}
O. M. Bulashenko, J. M. Rubi, and V. A. Kochelap, Phys, Rev, B {\bf 62}, 8184 (2000).

\bibitem{Gramespacher}
T. Gramespacher, and M. B\"uttiker, Phys. Rev. Lett. {\bf 81}, 2763 (1998).

\bibitem{Kuznetsov}
V. V. Kuznetsov, E. E. Mendez, J. D. Bruno, and J. T. Pham, Phys. Rev. B {\bf 58}, R10 159 (1998).

\bibitem{Simonian}
N. Simonian, J. B. Li and K. Likharev, Nanotechnology {\bf 18}, 424006 (2007).

\bibitem{Kudasov}
Y.B. Kudasov, Bulletin of the Russian Academy of Sciences: Physics {\bf 72}, 2 (2008).

\bibitem{Rakshit}
T. Rakshit, G. C. Liang, A. W. Ghosh, and Supriyo Datta, Nano Letters {\bf 4}, 10 (2004).

\bibitem{Song}
W. Song, E. E. Mendez, V. Kuznetsov, and B. Nielson, Appl. Phys. Lett {\bf 82}, 10 (2003).

\bibitem{Blanter2}
Y. M. Blanter and M. B\"uttiker, Phys. Rev. B {\bf 59}, 10217 (1999).

\bibitem{Lang}
N.D. Lang, Phys. Rev. B {\bf 55}, 9364 (1997); N.D. Lang and Ph. Avouris, Phys. Rev. Lett. {\bf 84}, 358 (2000).

\bibitem{chen}
Y.C. Chen and M. Di Ventra, Phys. Rev. B {\bf 67}, 153304 (2003).

\bibitem{Larade}
B. Larade, J. Taylor, H. Mehrez, H. Guo, Phys. Rev. B {\bf 64}, 075420 (2001).

\bibitem{but1}
M. Bu¨ttiker, Phys. Rev. B {\bf 46}, 12 485 (1992).

\bibitem{Wei}
Y. D. Wei, B. G. Wang and J. Wang, Phys. Rev. B {\bf 60}, 16900 (1999).

\bibitem{Taylor}
J. Taylor, H. Guo, and J. Wang, Phys. Rev. B {\bf 63}, 245407 (2001); {\bf 63} 121104 (2001).

\bibitem{Do}
V. N. Do, P. Dollfus and V. L. Nguyen, J. Appl. Phys. {\bf 100}, 093705 (2006).

\bibitem{Haug}
H. Haug and A. Jauho, Quantum Kinetics in Transport and Optics of Semiconductors (Springer, Berlin, 1998), Chap. 12.

\bibitem{Datta}
S. Datta, Electronic Transport in Mesoscopic Systems (Cambridge University Press, Cambridge, 1995), Chap. 8.

\bibitem{Langreth}
D. C. Langreth, in Linear and Nonlinear Electron Transport in Solids, ed. by J. T. Devreese and E. Van Doren (Plenum, New York, 1976).






\end{thebibliography}
\end{document}